\documentstyle[12pt,preprint,prc,aps,psfig,floats]{revtex}
\input psfig
\def\lsim{\mathrel{\rlap{\lower4pt\hbox{\hskip1pt$\sim$}}
    \raise1pt\hbox{$<$}}}         %less than or approx. symbol
\def\gsim{\mathrel{\rlap{\lower4pt\hbox{\hskip1pt$\sim$}}
    \raise1pt\hbox{$>$}}}         %greater than or approx. symbol
\begin{document}
\draft

\title{The luminosity constraint on solar neutrino fluxes}
\author{{John N. Bahcall}\thanks{jnb@ias.edu}} 
\address{School of Natural Sciences, 
Institute for Advanced Study, Princeton, NJ 08540}
\maketitle

\begin{abstract}
A specific linear combination of the total solar neutrino fluxes must
equal the measured solar photon luminosity if nuclear fusion reactions
among light elements are responsible for solar energy generation.
This luminosity constraint, previously used in a limited form in
testing the no neutrino oscillation hypothesis, is derived in a
generality that includes all of the relevant solar neutrino fluxes and
which is suitable for analyzing the results of many different solar neutrino
experiments. With or without allowing for neutrino oscillations, the
generalized luminosity constraint can be used in future analyses of
solar neutrino data.  Accurate numerical values for
the linear coefficients are provided.
\end{abstract}


\section{Introduction}
\label{sec:introduction}

If nuclear fusion reactions among light elements are responsible for
the solar luminosity, then a specific linear combination of solar
neutrino fluxes must equal the solar constant.  More explicitly, we
shall see that

\begin{equation}
{L_\odot\over 4\pi (A.U.)^2} = \sum\limits_i \alpha_i \Phi_i~, 
\label{eq:genconstraint}
\end{equation}
where $\L_\odot$ is the solar luminosity measured at the earth's
surface and 1 $A.U.$ is the average earth-sun distance.  The
coefficient $\alpha_i$ is the amount of energy provided to the star by
nuclear fusion reactions associated with each of the important solar
neutrino fluxes, $\Phi_i$.  Equation~(\ref{eq:genconstraint}) is known
as the luminosity constraint. The utility and power of the luminosity
constraint are due to the fact that the numerical values of the
$\alpha_i$'s are determined by the differences between nuclear masses
and are independent of details of the solar model to an accuracy of
one part in $10^4$ or better.

The conservation laws of baryon number,
lepton number, and charge are the basis for the luminosity constraint
and are contained within the reaction equation that describes the
fusion of hydrogen to make helium within stars,
\begin{equation}
4 p ~\longrightarrow ~ ^4{\rm He} ~+~ e^+ ~+~ 2\nu_e ~+~{\rm (thermal ~
energy)}. 
\label{eq:conservation}
\end{equation}
The amount of thermal energy that is delivered to the star depends
upon the particular set of reactions that occur in realizing
Eq.~(\ref{eq:conservation}). If, for example, the relation shown in
Eq.~(\ref{eq:conservation}) involves the production of the relatively
high energy ${\rm ^8B}$ neutrinos, then less thermal energy is
communicated to the star than if only low energy $pp$ neutrinos are
produced.

The sum in Eq.~(\ref{eq:genconstraint}) should be taken over all the
neutrino fluxes whose associated nuclear fusion reactions could in
principle contribute significantly to the energy budget of the sun.
If no detailed knowledge derived from solar models is used to limit
the appearance of terms in Eq.~(\ref{eq:genconstraint}), then the
luminosity constraint can be applied to tests of the hypothesis of no
neutrino oscillations. The luminosity constraint provides an
additional condition that must be satisfied by neutrino fluxes that
are otherwise allowed to have arbitrary amplitudes when fit to the
available solar neutrino
data~\cite{spiro90,dar91,hata94,castellani94,berezinsky94,parke95,fogli95,heeger96,bahcallkrastev96,bks98}.

\subsection{Previous work}
\label{subsec:previous}

Spiro and Vignaud \cite{spiro90} first proposed, in a lucid and
insightful paper, the use of the luminosity constraint as a test of
the null hypothesis for solar neutrino propagation, i.e. as a test
independent of solar models of the assumption of no neutrino
oscillations.  Following these authors, most of the pioneering
applications of the luminosity constraint (see, e.g.,
~\cite{dar91,hata94,castellani94,berezinsky94,parke95,heeger96}) have
approximated the solar neutrino spectrum by grouping the neutrinos
into three sets, the low-energy (principally $pp$ )neutrinos,
intermediate energy neutrinos (usually taken to be either ${\rm ^7Be}$
neutrinos only or the ${\rm ^7Be}$ neutrinos plus the CNO and $pep$
neutrinos), and finally the high-energy (${\rm ^8B}$) neutrinos. This
approximation was necessary and appropriate when the number of solar
neutrino experiments was small (two or four), but is no longer
required or optimal now that the number of solar neutrino experiments
is six (chlorine \cite{chlorine}, Kamiokande \cite{kamiokande}, SAGE
\cite{sage}, GALLEX + GNO~\cite{gallex,gallexgno}, Super-Kamiokande
\cite{superk}, and SNO \cite{sno}) and growing
(BOREXINO~\cite{borexino} and ICARUS~\cite{icarus}), with additional
experiments in the planning stages.

The derivation given here explains
(and, in some cases, corrects) the the results that were simply stated
in Ref.~\cite{bahcallkrastev96}.

All previous discussions with which I am familiar, including
Ref.~\cite{bahcallkrastev96}, have implemented the luminosity
constraint in the context of showing  that solar fluxes with arbitrary
amplitudes, but without the distortion of the energy spectrum implied
by neutrino oscillations, do not fit the available data on the measured
rates of solar neutrino experiments.  The papers by Hata, Bludman, and
Langacker~\cite{hata94}, Parke~\cite{parke95}, Heeger and
Robertson~\cite{heeger96}, and Bahcall, Krastev, and
Smirnov~\cite{bks98} were important in persuading many
physicists who are not familiar with solar models that a particle
physics solution was required for the solar neutrino problem.

\subsection{What is this paper about?}
\label{subsec:aboutwhat}

My goal in this paper is to provide a general formulation for the
luminosity constraint that can be used in future analyses, with or without
allowing for neutrino oscillations, that may include six or more
experiments. I also want to provide a specific derivation, lacking in
the literature, for the coefficients in
Eq.~(\ref{eq:genconstraint}). The lack of a general derivation in the
literature has led to a confusion about the basis for the luminosity
constraint and to significant errors in the published values of some
coefficients.

In this paper, I derive the coefficients for the luminosity constraint
for all seven of the important neutrino fluxes shown in
Table~\ref{tab:nuconstraint}.  After deriving the coefficients for the
general form of the luminosity relation, I discuss the most
appropriate approximations to make in analyzing data sets in which the
number of measured neutrino event rates is not sufficient to allow a
statistically meaningful application of the full luminosity
constraint.  The discussion in this paper is a natural generalization
of the treatment given in the very important paper by Hata {\it et
al.} \cite{hata94}, which presented cogently the argument that the
measured solar neutrino event rates (chlorine`\cite{chlorine},
Kamiokande~\cite{kamiokande}, SAGE~\cite{sage}, and GALLEX~\cite{gallex})
 required new physics, even before the epochal Super-Kamiokande~\cite{superk}
and SNO~\cite{sno} measurements.

The formulations presented in previous discussions can be recovered
from the present analysis by assuming that one or all of the following
assumptions is valid: i) certain neutrino fluxes are zero; ii) the CNO
neutrino fluxes (from ${\rm ^{13}N}$ and ${\rm ^{15}O}$ beta-decay)
are equal; or iii) the standard solar model ratio of $pep$ neutrino
flux to $pp$ neutrino flux is correct. 

In the future, the generalized luminosity constraint can and should be
implemented in analyses that determine solar neutrino parameters. The
additional constraint provided by the measured solar luminosity 
will be especially important when $pp$ and ${\rm ^7Be}$ neutrino
fluxes are measured as well as the ${\rm ^8B}$ neutrino flux. As more
experimental data become available, the 
analyses of neutrino oscillations will become more independent of the
standard solar model and it will be natural and convenient to
incorporate the luminosity constraint, Eq.~(\ref{eq:genconstraint}).

The luminosity constraint can be written conveniently in a
dimensionless form by considering the ratios of all the neutrino fluxes
to the values predicted by the standard solar model, and dividing both
sides of Eq.~(\ref{eq:genconstraint}) by $L_\odot/[4\pi(A.U.)^2]$. We
obtain

\begin{equation}
 1 = \sum\limits_i \left({\alpha_i\over 10~ {\rm MeV}}\right) a_i
 \phi_i~,
\label{eq:dimensionless}
\end{equation}
where the 
dimensionless neutrino fluxes, $\phi_i$, are the ratios of the
true neutrino fluxes to the neutrino fluxes predicted by the BP2000
standard solar model~\cite{bp2000}, i.e., 
\begin{equation}
\phi_i \equiv \Phi_i/\Phi_i {\rm (BP2000)}~.
\label{eq:phii}
\end{equation}
The 
quantities $a_i$ are the ratios of $\Phi_i {\rm (BP2000)}$
to the characteristic solar photon flux defined by
$L_\odot/[4\pi(A.U.)^2({\rm 10 Mev})]$. Thus
\begin{equation}
a_i \equiv \Phi_i {\rm (BP2000)}/\left(8.5272 \times 10^{10} 
{\rm cm^{-2}~s^{-1}}\right)~.
\label{eq:ai}
\end{equation}
In calculating the characteristic solar flux, I have used the recent
best-estimate solar luminosity (see Ref.~\cite{solarlum}), $3.842
\times 10^{33} {\rm ~erg s^{-1}}$, that is derived from all the
available satellite data. 

Depending on the context, I shall use $\phi_i$ to refer to either the
neutrino flux produced locally or integrated over the entire sun. This
dual usage will not cause any confusion since the specific meaning of
$\phi_i$ will be clear in all cases. Moreover, because of the
linearity of the averaging process, equations that are valid locally
have the same form when the results are integrated over the entire sun.

Table~\ref{tab:nuconstraint} provides the numerical values for the
dimensionless form of the luminosity constraint.

Some people may prefer to use the luminosity constraint in its
dimensional form (Eq.~\ref{eq:genconstraint}).  In this case, one can
ignore the values of $a_i$ given in Table~\ref{tab:nuconstraint} and
use just the tabulated values of $\alpha_i$. If one wants to use the
constraint in its dimensionless form but with a different basis set of
neutrino fluxes instead of those from BP2000, then one can replace the
$a_i$ given here by new values computed from Eq.~(\ref{eq:ai}) with the
different model fluxes.

The linear equality of Eq.~(\ref{eq:dimensionless}) must be
supplemented by the physical requirement that the number of nuclear
reactions that terminate the proton-proton chain not exceed the number
of initiating nuclear reactions, which can be translated into the
following inequality~\cite{bahcallkrastev96},
\begin{equation}
\phi \left({\rm ^7Be}\right) + \phi \left({\rm ^8B}\right)\  \leq\  \phi
({\rm pp}) + \phi ({\rm pep})~.
\label{addconstraint}
\end{equation}
The physical basis of Eq.~(\ref{addconstraint}) is that the ${\rm
^3He}$ nuclei, which ultimately give rise to ${\rm ^7Be}$ and ${\rm
^8B}$ neutrinos via the nuclear reaction ${\rm ^3He} (\alpha,
\gamma){\rm ^7Be}$, are created by pp and pep\ reactions.  One pp or
pep\ reaction must occur in order to supply the ${\rm ^3He}$ nucleus
that is burned each time a ${\rm ^7Be}$ or ${\rm ^8B}$ neutrino is
produced.  In principle, Eq.~(\ref{eq:dimensionless}) (or
Eq.~\ref{eq:genconstraint}) considered separately permits a ${\rm
^7Be}$ neutrino flux that is twice as large as is allowed by
Eq.~(\ref{addconstraint}).  Since the ${\rm ^{14}N} (p, \gamma){\rm
^{15}O}$ reaction is the slowest process in the CNO cycle, one must
also have~\cite{bahcallkrastev96}
\begin{equation}
\phi\left({\rm ^{15}O}\right) \ \leq\  \phi \left({\rm ^{13}N}\right)~.
\label{CNOineq}
\end{equation}
With the currently available solar neutrino data, the additional
constraints provided by Eq.~(\ref{addconstraint}) and
Eq.~(\ref{CNOineq}) are automatically satisfied by the best-fit
solutions for the undistorted neutrino fluxes to the measured event
rates~\cite{bahcall01}.

\begin{table}[!t]
\tightenlines
\caption[]{\label{tab:nuconstraint} {\bf Luminosity constraint:
neutrino characteristics.} The average neutrino energies are taken
from Ref.~\cite{bahcall94} for ${\rm ^7Be}$ and from
Ref.~\cite{bahcall97} for all other sources.  The neutrino energies
include thermal effects from electron and ion motion and from the
solar temperature profile, as well as atomic ionization effects.  The
nuclear data are taken from \cite{firestone96}. The quantities
$\alpha$ and $a$ are defined in
Eqs.~(\ref{eq:dimensionless})--(\ref{eq:ai}), the dimensionless
form of the luminosity constraint.}
\begin{tabular}{lllcc} 
\noalign{\medskip}
\multicolumn{1}{c}{Flux}&\multicolumn{1}{c}{Reaction}
&$\langle E_\nu\rangle_\odot$&$\alpha$&$a$\\
&&(MeV)&(10 MeV)\\
\noalign{\medskip}
\hline
\noalign{\medskip}
$\phi (pp)$&$p + p \to {\rm ^2H} + e^+ + \nu_e$&0.2668&1.30987&0.6978\\
$\phi (pep)$&$p + e^- + p \to {\rm ^2H} +
\nu_e$&1.445&1.19193&0.001642\\
$\phi (hep)$&${\rm ^3He} + p \to {\rm ^4He} + e^+ \nu_e$&9.628&0.37370&1.09E-07\\
$\phi ({\rm ^7Be})$&${\rm ^7Be} + e^- \to {\rm ^7Li} +
\nu_e$&0.814\tablenotemark[1]&1.26008&0.05594\\
$\phi ({\rm ^8B})$&${\rm ^8B} \to {\rm ^8Be} + e^+ +
\nu_e$&6.735&0.66305&0.0000592\\
$\phi ({\rm ^{13}N})$&${\rm ^{13}N} \to {\rm ^{13}C} + e^+ +
\nu_e$&0.706&0.34577&0.006426\\
$\phi ({\rm ^{15}O})$&${\rm ^{15}O} \to {\rm ^{15}N} + e^+ + \nu_e$&0.996&2.15706&0.005629\\
\noalign{\medskip}
\end{tabular}
\tablenotetext[1]{89.7\% 0.8631 MeV and 10.3\% 0.3855 MeV.}
\end{table}

In Sec.~\ref{sec:derivation}, I derive explicit expressions for the
$\alpha$ coefficients that appear in Eq.~(\ref{eq:dimensionless}) in
terms of measured atomic mass differences and computed neutrino energy
losses. I summarize in Sec.~\ref{sec:assumptions} the principal
assumptions that are used in the derivation and discuss the results
and application strategies in Sec.~\ref{sec:discussion}.

\section{Derivation of the luminosity constraint}
\label{sec:derivation}

In this section, I shall derive expressions for the energy
coefficients, $\alpha_i$, that appear in the luminosity constraint,
Eq.~(\ref{eq:genconstraint}). Section~\ref{subsec:cnonus} treats the
simpler case of the CNO neutrinos and Sec.~\ref{subsec:ppnus} derives
the coefficients for neutrino fluxes produced by reactions in the $pp$
chain. The numerical values given in Table~\ref{tab:nuconstraint} were
calculated using Eqs. (\ref{eq:alphan13}) and (\ref{eq:alphao15}) of
Sec.~\ref{subsec:cnonus} and
Eqs.~(\ref{eq:lumepsilon})--(\ref{eq:alphaepsilon}) of
Sec.~\ref{subsec:ppnus}.

The coefficients $\alpha_i$ that appear in the luminosity constraint
 are independent of details of the solar model to high accuracy. As we
 shall see below, the $\alpha's$ represent the differences between
 nuclear masses with only small corrections, included here, due to the
 average thermal energy of the fusing particles. The largest of the
 thermal corrections amounts to $0.7$\% of the average neutrino energy
 loss for the $pp$ neutrinos~\cite{bahcall97} and $0.15$\% for the
 ${\rm ^7Be}$~\cite{bahcall94} neutrinos. For all other solar neutrino
 sources, the thermal corrections to the energy loss are much
 smaller~\cite{bahcall91}. Since the total neutrino energy loss is itself
 only a small fraction of the $\alpha$'s we compute in the following
 two subsections, one can show that thermal energy affects the
 values of $\alpha$ given in Table~\ref{tab:nuconstraint} by less than
 or of order $0.01$\% for all of the important solar neutrino energy
 sources.

In carrying out the calculations, we  will use $R_{ij}$ to represent
the reaction rate per unit of time per unit of
volume between two fusing ions, $i$ and $j$, where 
\begin{equation}
R_{ij} = {\langle i,j\rangle n(i)n(j)\over (1 + \delta_{ij})}~.
\label{eq:rij}
\end{equation}
Here $\langle i, j\rangle$ is the local thermal average of $\sigma v$,
the product of the relative velocity of particles $i,j$ and their
interaction cross section, and the Kronecher delta prevents double
counting of identical particles.  For example, $R_{34} = \langle {\rm
^3He, ^4He}\rangle n\left({\rm ^3He}\right) n\left({\rm ^4He}\right)$
and $R_{pp} = \langle p,p\rangle n(p)^2/2$.

In the following, 
I shall denote the average neutrino energy from a particular
nuclear reaction, $X$, by $<E_\nu>(X)$.  
Table~\ref{tab:nuconstraint} lists accurate values for these neutrino
energies. The values given in Table~\ref{tab:nuconstraint}
are averaged over the energy spectrum from each source and include
corrections for solar effects such as contributions from the thermal
motion of the fusing ions, averages over ionization states, and the
temperature profile of the sun.  The average solar neutrino energy
losses are taken from Refs.~\cite{bahcall94,bahcall97}. The masses,
for example $M\left({\rm ^{13}C}\right)$ or $M\left({\rm
^{1}H}\right)$, that appear in the equations of
Sec.~\ref{subsec:cnonus} and Sec.~\ref{subsec:ppnus} are atomic
masses which I have taken from Ref.~\cite{firestone96}.

For convenience in the calculations, we will introduce a fictitious
neutrino flux density,
\begin{equation}
\Phi_{33} \equiv \langle{\rm ^3He, ^3He}\rangle n \langle{\rm ^3He,
^3He}\rangle/2~,
\label{eq:33flux}
\end{equation}
which is the rate of the ${\rm ^3He}({\rm ^3He},2p){\rm ^4He}$
reaction (and would be the flux density of neutrinos produced by this
reaction if the ${\rm ^3He}-{\rm ^3He}$ reaction gave rise to neutrinos). This
fictitious flux will not appear in any of the final formulae.

\subsection{CNO neutrinos}
\label{subsec:cnonus}

The ${\rm ^{13}N}$ beta-decay corresponds to the thermal energy
derived from the reactions ${\rm ^{12}C} (p,\gamma){\rm ^{13}N}$ and 
${\rm ^{13}N} \to {\rm ^{13}C} + e^+ + \nu_e$. Hence
\begin{equation}
\alpha \left({\rm ^{13}N}\right) = M\left({\rm ^{12}C}\right) + M
\left({\rm ^1H}\right) - M\left({\rm ^{13}C}\right)- \langle
E_\nu\rangle \left({\rm ^{13}N}\right)~.
\label{eq:alphan13}
\end{equation}

Similarly, the energy associated with the ${\rm ^{15}O}$ beta-decay
derives from the reactions ${\rm ^{13}C} (p,\gamma){\rm ^{14}N}$,
${\rm ^{14}N} (p,\gamma){\rm ^{15}O}$, ${\rm ^{15}O} \to {\rm ^{15}N}+
e^+ + \nu_e$, and ${\rm ^{15}N}(p,\alpha){\rm ^{12}C}$. Therefore,
\begin{equation}
\alpha \left({\rm ^{15}O}\right) = 3M \left({\rm ^1H}\right) +
M\left({\rm ^{13}C}\right) - M\left({\rm ^4He}\right) - M\left({\rm
^{12}C}\right) - \langle E_\nu\rangle \left({\rm^{15}O}\right)~.
\label{eq:alphao15}
\end{equation}

The neutrino flux from ${\rm ^{17}F}$ beta-decay is a potential
measure of the primordial ${\rm ^{16}O}$ abundance in the sun (see
Ref.~\cite{bahcalletal82}), but does not play a significant role in
the generation of the solar luminosity.  For completeness, I include
here the coefficients $\alpha$ and $a$ that describe the reactions
${\rm ^{16}O} (p,\gamma){\rm ^{17}F}$ and ${\rm ^{17}F} \to {\rm
^{17}O}+ e^+ + \nu_e$.  The appropriate linear coefficients for use in
Eq.~(\ref{eq:dimensionless}) are
\begin{equation}
\alpha\left({\rm ^{17}F}\right) = 2.363~{\rm MeV}~,\,\,\,\,\,
a\left({\rm ^{17}F}\right) = 1.09E-07~.
\label{eq:alphaf17}
\end{equation}
The value of $a\left({\rm ^{17}F}\right)$ is so small (because of the
high Coulomb barrier for this reaction) that the flux $\phi({\rm ^{17}F})$ is
not relevant for practical applications of the luminosity constraint.

\subsection{$pp$ neutrinos}
\label{subsec:ppnus}

The analysis of reactions in the $pp$ chain is simplified by the
fact that ${\rm ^2H}$ and ${\rm ^3He}$ are burned quickly at solar
temperatures \cite{book}.  The lifetime for nuclear burning of ${\rm ^2H}$ is
$\sim 10^{-8}$ yr and the lifetime of ${\rm ^3He}$ is $\sim 10^5$ yr.
These values are both very small compared to the $10^{10}$~yr lifetime
of a proton (which is destroyed primarily by the $pp$ reaction).
Therefore, it is an excellent approximation to assume that both ${\rm
^2H}$ and ${\rm ^3He}$ are in local kinetic equilibrium (rate of
destruction equals rate of production).

From the local equilibrium of ${\rm ^2H}$, $dn({\rm ^2H})/dt = 0$, one
has
\begin{equation}
\langle{\rm ^1H, ^1H}\rangle n \left({\rm ^1H}\right)^2/2~+~
\langle{\rm ^1H}, e^- + {\rm H}\rangle n (e) n({\rm H}) ~=~ \langle {\rm
^1H, ^2H}\rangle n\left({\rm ^1H}\right)n\left({\rm ^2H}\right)~.
\label{eq:h2equilib}
\end{equation}
Equation~(\ref{eq:h2equilib}) states that the production of deuterium
via the $pp$ and $pep$ reactions is balanced by the destruction of
deuterium via the ${\rm ^2H}(p,\gamma){\rm ^3He}$ reaction.  The
equilibrium of ${\rm ^3He}$, $dn{\rm (^3He)}/dt = 0$, implies

\begin{eqnarray}
&&\langle{\rm ^1H, ^2H}\rangle n \left({\rm ^1H}\right)n\left({\rm
^2H}\right) = \langle{\rm ^3He, ^3He}\rangle n(3)^2\nonumber +
\langle{\rm ^3He, ^4He}\rangle n\left({\rm ^3He}\right) n\left({\rm
^4He}\right) \\ &&  ~~~~~~~~~~~~~~~~~~~~~~~~~~~~~~+ \langle{\rm ^3He,
^1H}\rangle n \left({\rm ^3He}\right) n\left({\rm ^1H}\right)~.
\label{eq:he3equilib}
\end{eqnarray}
Equation~(\ref{eq:he3equilib}) describes the fact that the rate
of production of ${\rm ^3He}$ via the ${\rm ^2H + ^1H}$ reaction is balanced by
the destruction of ${\rm ^3He}$ via the ${\rm ^3He}-{\rm ^3He}$, ${\rm
^3He}$-${\rm ^4He}$, and
$hep$ reactions.  For the term describing the ${\rm ^3He}$-${\rm ^3He}$
reaction in Eq.~(\ref{eq:he3equilib}), the factor of one-half from the
identity of the fusing particles is cancelled by the factor of two
representing the destruction of two ${\rm ^3He}$ ions.

Combining Eqs.(\ref{eq:33flux}), (\ref{eq:h2equilib}), and
(\ref{eq:he3equilib}), we find

\begin{equation}
\Phi (p\hbox{-}p) + \Phi (pep) = 2 \times \Phi (3,3) + \Phi (hep) + \Phi
\left({\rm ^7Be}\right) + \Phi\left({\rm ^8B}\right)~.
\label{eq:ppidentity}
\end{equation}

Let $\epsilon_i$ represent the thermal energy released to the star as
a result of the nuclear fusion reactions associated directly with each
neutrino producing reaction. Then we can write

\begin{eqnarray}
&&L_\odot/\left[4\pi (A.U.)^2\right] = {\rm CNO~terms} + \epsilon_{pp}
\Phi (p\hbox{-}p) + \epsilon_{pep} \Phi (pep)\nonumber\\
&&~~~~~~~~~~~~~~~~~~~~~+
\left(\epsilon_{33}/2\right)\left[\Phi(p\hbox{-}p) + \Phi (pep) - \Phi
(hep) - \Phi\left({\rm ^7Be}\right) - \Phi \left({\rm
^8B}\right)\right]\nonumber\\ &&~~~~~~~~~~~~~~~~~~~~~+\epsilon_{hep}
\Phi (hep) + \left(\epsilon_{34} + \epsilon_{e7}\right) \Phi
\left({\rm ^7Be}\right) + \left(\epsilon_{34} +
\epsilon_{p,7}\right)\Phi\left({\rm ^8B}\right)~.
\label{eq:lumepsilon}
\end{eqnarray}
I have used Eq.~(\ref{eq:ppidentity}) to eliminate the fictitious flux
$\phi (3,3)$ from Eq.~(\ref{eq:lumepsilon}).  This
substitution associates with each of the real neutrino fluxes in the
$pp$ chain an additional energy contribution proportional to
$\epsilon_{33}$, the energy released to the star via the ${\rm
^3He}$-${\rm ^3He}$ reaction. Physically, these terms proportional to
$\epsilon_{33}$ represent a way of keeping track of how much energy
from the the ${\rm ^3He}$-${\rm ^3He}$ reaction should be associated
with the other neutrino fluxes.

The values of $\epsilon_i$ can be calculated by writing explicitly the
reaction equations that are associated with the production of each
neutrino flux. One finds

\begin{equation}
\epsilon_{pp} = 3M \left({\rm ^1H}\right) - M\left({\rm ^3He}\right) -
\langle E_\nu\rangle (p\hbox{-}p)~,
\label{eq:epp}
\end{equation}

\begin{equation}
\epsilon_{pep} = 3M \left({\rm ^1H}\right) - M\left({\rm ^3He}\right)
- \langle E_\nu\rangle(pep)~,
\label{eq:epep}
\end{equation}

\begin{equation}
\epsilon_{33} = 2 M\left({\rm ^3He}\right) - M\left({\rm ^4He}\right)
- 2 M\left({\rm ^1H}\right)~,
\label{eq:e33}
\end{equation}

\begin{equation}
\epsilon_{34} =  M\left({\rm ^3He}\right) + M\left({\rm ^4He}\right)
- M\left({\rm ^7Be}\right)~,
\label{eq:e34}
\end{equation}

\begin{equation}
\epsilon_{e7} = M \left({\rm ^7Be}\right) + M\left({\rm ^1H}\right) -
2 M\left({\rm ^4He}\right) - \langle E_\nu\rangle \left({\rm
^7Be}\right)~,
\label{eq:ee7}
\end{equation}

\begin{equation}
\epsilon_{p,7} = M\left({\rm ^7Be}\right) + M\left({\rm ^1H}\right) -
2M\left({\rm ^4He}\right) - \langle E_\nu\rangle \left({\rm ^8B}\right)~,
\label{eq:ep7}
\end{equation}
and 
\begin{equation}
\epsilon_{hep} = M\left({\rm ^3He}\right) + M \left({\rm ^1H}\right) -
M\left({\rm ^4He}\right) - \langle E_\nu\rangle (hep)~.
\label{eq:ehep}
\end{equation}
In calculating $\epsilon_{e7}$, one must average over the two ${\rm
^7Be}$ neutrino lines with the appropriate weighting and include the
$\gamma$-ray energy from the 10.3\% of the decays that go to the first
excited state of ${\rm ^7Li}$.

The values of the $\alpha$'s can be determined using the following
relations between the $\alpha$ coefficients and the $\epsilon$
coefficients that follow from Eq.~(\ref{eq:lumepsilon}).
We have 
\begin{eqnarray}
&& \alpha (p\hbox{-}p) = \epsilon_{pp} + 0.5 \epsilon_{33},\,\,\,\,
\alpha (pep) = \epsilon_{pep} + 0.5 \epsilon_{33},\,\,\,\, \alpha
\left({\rm ^7Be}\right) = \epsilon_{34} + \epsilon_{e7} - 0.5
\epsilon_{33}, \nonumber \\ && \alpha\left({\rm ^8B}\right) =
\epsilon_{34} + \epsilon_{p,7} - 0.5 \epsilon_{33}, \, \,\,\, \alpha
(hep) = \epsilon_{hep} - 0.5 \epsilon_{33}~.
\label{eq:alphaepsilon}
\end{eqnarray}

\section{What assumptions are made in deriving the luminosity
constraint?}
\label{sec:assumptions}

The basic assumption made in deriving the luminosity constraint is
that nuclear fusion reactions among light elements are responsible for
the observed solar luminosity. More specifically, I assume in
Sec.~\ref{sec:derivation} that the specific nuclear reactions that
have been recognized~\cite{book,bethe,parker,clayton,fowler} over the
six decades since Hans Bethe's epochal work on the subject as being
most important at temperatures of order a keV are indeed the fusion
reactions that power the sun. The characteristic temperature of the
sun can be estimated relatively well without making use of a detailed
model~\cite{book,bethe}.

In order to derive Eq.~(\ref{eq:h2equilib}) and
Eq.~(\ref{eq:he3equilib}), it is necessary to assume that both $^2$H
and $^3$He are in local kinetic equilibrium (rate of creation equal
rate of destruction). As discussed in Sec.~\ref{subsec:ppnus}, this
is an excellent approximation because the lifetimes for nuclear
burning of these isotopes are short compared to the evolutionary time
scale for the sun.

\section{Discussion}
\label{sec:discussion}

I have given in Sec.~\ref{sec:derivation} an explicit derivation of
the luminosity constraint and have presented in
Table~\ref{tab:nuconstraint} coefficients that can be used in the
dimensionless form of the constraint, Eq.~(\ref{eq:dimensionless}).
The coefficients that are given in Table~\ref{tab:nuconstraint} are
calculated accurately and include small corrections to the neutrino
energy release that result from the high temperatures ($\sim {\rm
keV}$) in the region in which fusion reactions occur in the sun.
The thermal corrections to the energy release are less than or of
order $0.01$\% in all cases. 

The form of the luminosity constraint given here includes all the
important solar neutrino fluxes.  One can recover approximately the
coefficients used by previous authors who have combined neutrino
fluxes, or who have considered only a reduced set of fluxes, by making
the relevant choices among the fluxes listed in
Table~\ref{tab:nuconstraint}. However, the reader is warned not to
expect precise agreement; there are many inaccurate numerical values
in the published papers.\footnote{In fact, I am responsible for an
egregious but unimportant error. In Ref.~\cite{bahcallkrastev96}, the
listed value for $\alpha (hep)$ is wrong because I neglected
to subtract the value of $\epsilon_{33}/2 = 6.4298$ MeV in going from
$\epsilon_{hep}$ to $\alpha (hep)$ [see
Eq.~(\ref{eq:alphaepsilon}) of the present paper].  The error is
unimportant, although embarrassing, because the Super-Kamiokande
experiment has placed \cite{superk} a strong upper limit on the $hep$
flux; this upper limit  is also in agreement with the 
predicted $hep$ flux for the standard solar model~\cite{bp2000}. }

The generalized form of the luminosity constraint presented here can
be used, as the more restricted constraint has been used in the past,
to help test the validity of the null hypothesis for solar neutrino
oscillations. In this test
(cf. Refs.~\cite{spiro90,dar91,hata94,castellani94,berezinsky94,parke95,fogli95,heeger96,bahcallkrastev96}),
the neutrino fluxes are allowed to have arbitrary amplitudes subject
to Eq.~(\ref{eq:dimensionless}) and the condition that the energy
spectra are undistorted by neutrino oscillations.  One should also
impose the two inequalities, Eq.~(\ref{addconstraint}) and
Eq.~(\ref{CNOineq}), that follow from the requirement that the number
of nuclear fusion reactions that terminate the proton-proton chain and
the CNO cycle not exceed the number of inititiating nuclear reactions.

In the past, applications of the luminosity constraint have been
limited to tests of the no oscillation hypothesis.  Nothing in the
derivation of the luminosity constraint (or the supplementary
inequalities, Eq.~\ref{addconstraint} and Eq.~\ref{CNOineq}), requires
this limitation in the range of applications.  In the future, when
more experimental data are available, the luminosity constraint can be
used together with the measured solar neutrino interaction rates,
energy spectra, and time dependences, to help determine neutrino
oscillation parameters.

\acknowledgments I am grateful to A. Friedland for a valuable
discussion. This research is supported by NSF Grant No. PHY0070928.

\end{document}